\documentclass[prc,nofootinbib,twocolumn,showpacs]{revtex4-1}
\usepackage{amssymb,amsmath}
\usepackage{epsfig,psfrag}

\usepackage{mathrsfs}
\usepackage{amsfonts}

\begin{document}
%JPV
\title{Deuteron-equivalent and phase-equivalent interactions within light nuclei}
%\title{Deuteron-equivalent phase-equivalent transformation and its manifestation in few-body nuclei}
\author{A. M. Shirokov}
\affiliation{Skobeltsyn Institute of Nuclear Physics, Moscow State
University, Moscow 119991, Russia}
\affiliation{Department of Physics and Astronomy, Iowa State University, Ames, IA 50011, USA}
\affiliation{Pacific National University, 136
Tikhookeanskaya Street, Khabarovsk 680035, Russia}
\author{V. A. Kulikov}
\affiliation{Skobeltsyn Institute of Nuclear Physics, Moscow State
University, Moscow 119991, Russia}
\affiliation{Pacific National University, 136
Tikhookeanskaya Street, Khabarovsk 680035, Russia}
%\affiliation{Department of Physics and Astronomy, Iowa State University, Ames, IA 50011, USA}
\author{A. I. Mazur}
\affiliation{Pacific National University, 136
Tikhookeanskaya Street, Khabarovsk 680035, Russia}
\author{J. P. Vary}
\affiliation{Department of Physics and Astronomy, Iowa State University, Ames, IA 50011, USA}
\author{P. Maris}
\affiliation{Department of Physics and Astronomy, Iowa State University, Ames, IA 50011, USA}

\begin{abstract}
\begin{description}
\item[Background] Phase-equivalent transformations (PETs) are well-known in quantum scattering 
and inverse scattering theory. PETs do not affect scattering phase shifts and bound state energies
of two-body system but 
are conventionally supposed to modify  two-body bound state 
%properties
%usually modify other 
observables such as the rms radius and electromagnetic moments.
\item[Purpose] In order to preserve all bound state observables, we propose a new 
particular case of PETs, a deuteron-equivalent
transformation (DET-PET), which leaves unchanged not only scattering phase shifts and 
bound state (deuteron) binding energy but also the bound state wave function. 
\item[Methods] The construction of DET-PET is discussed; equations defining the
simplest DET-PETs are derived. We apply these simplest DET-PETs to the 
JISP16 $NN$ interaction and use the 
%obtained 
transformed $NN$ interactions in
calculations  of $^3$H and $^4$He binding energies in the No-core Full Configuration (NCFC) 
approach based on extrapolations of the No-core Shell Model (NCSM) basis space results 
to the infinite basis space.
\item[Results] We demonstrate the DET-PET modification of the $np$ scattering wave functions
%that are limited to regions near the nucleus 
and study the DET-PET manifestation  in the binding energies of $^3$H and $^4$He nuclei and 
their correlation (Tjon line).
\item[Conclusions] It is shown that some DET-PETs generate modifications of the central 
component while the others  modify the tensor  component of the $NN$ interaction. DET-PETs
are able to modify significantly the $np$ scattering wave functions 
%in regions near the nucleus
and hence the off-shell 
properties of the $NN$ interaction.  DET-PETs give rise to 
%essential variations of 
significant changes in
the binding energies of $^3$H (in the range of approximately 1.5 MeV) and $^4$He
(in the range of more than 9 MeV) and are able to modify the correlation patterns of 
binding energies of these nuclei.
\end{description}
 \end{abstract}
% 03.65.Nk	Scattering theory
% 21.30.-x	Nuclear forces
%21.10.Dr	Binding energies and masses
%21.45.Bc	Two-nucleon system
%21.60.De	Ab initio methods
%21.45.-v	 	Few-body systems

\maketitle

\section{Introduction}
Phase-equivalent transformations (PETs) of two-body interactions
 are well-known in quantum scattering theory \cite{Newton}. 
 PETs play an important role in the inverse scattering theory giving rise
to ambiguities in the derived potentials. Currently, there is intensive research on 
supersymmetric transformations (see reviews 
\cite{SUSY-review, SUSY-inversereview}) which are a particular case of PETs \cite{ShiSid}
associated with removing or adding bound states to the system. 

More traditional PETs
which do not change the on-shell properties of the two-body interaction, i.\,e. two-body 
scattering phase shifts and the energies
of the two-body discrete spectrum states, but modify the interaction off-shell,   were used
to study manifestations of off-shell properties of two-nucleon interactions
in many-nucleon systems. For example, a correlation between the nuclear matter binding
energy and its equilibrium density (the so-called Coester line) was studied with
phase-equivalent $NN$ interactions 
in Ref. \cite{Coester}. PETs were used to modify the nucleon-cluster interaction in order to
obtain a correct description of the nuclear binding energies in cluster model studies of
Ref. \cite{PHT,LurAnn}. Various versions of the realistic JISP $NN$ interaction (JISP6 \cite{JISP6} and
JISP16 \cite{JISP16,JISP16_web}) were obtained by means of PETs applied to the initial ISTP $NN$
interaction \cite{ISTP} obtained in the $J$-matrix inverse scattering approach with 
the aim of improving
the description of binding energies of many-nucleon systems. 
The interaction JISP16
\cite{JISP16,JISP16_web} provides an accurate 
description of light nuclei 
 \cite{JISP16,YaF2008,Rila,JMPE2008, fbBonn, Izv2011,NCFC,LIT-JISP,VainBarnGaz, Slaus2007, OrlBarnLeid}
and was used to predict the
binding energy and spectrum of the exotic $^{14}$F nucleus \cite{14F} which were
confirmed later in the first experimental observation of this isotope \cite{14Fexp}.

We propose here a new type of PET, a deuteron-equivalent transformation (DET-PET).
Contrary to conventional PETs resulting in the modification of bound state and scattering wave
functions \cite{Newton, ShiSid,Coester,ISTP}, 
DET-PET guarantees that the transformed interaction generates not only the same
scattering phase shifts and two-body binding energy (or, more generally, bound state energies) 
but also the same bound state (deuteron) wave 
function as the initial untransformed interaction. 
The same method easily generalizes to preserve a set of bound state
wave functions.
DET-PET has the advantage of preserving the deuteron ground state observables.
On the other hand, DET-PET,
as well as any PET, modifies a two-body interaction off-shell, and hence manifests itself in 
many-body systems. 

One may naturally inquire whether PETs may lead to a better understanding of the appropriate 
off-shell behavior for the $NN$ interaction.  We note that what is appropriate depends on 
the adopted theoretical framework for the $NN$ interaction.  Since the interaction is not 
an observable, all approaches (meson exchange, EFT, lattice gauge, inverse scattering, ...) build 
in model assumptions (e. g., form factors, regulators, cutoffs, ...). Given those model assumptions, 
there are additional unexplored off-shell freedoms and we will show below how to explore those 
freedoms with constraints tied to $NN$ bound state observables.

After introducing the formulas defining DET-PET, we apply DET-PET to the JISP16
$NN$ interaction and illustrate various versions of DET-PET by respective modifications of
scattering wave functions at a few values of the DET-PET continuous parameter. A DET-PET
manifestation in many-body systems is illustrated by the study of binding energies of
$^3$H and $^4$He binding energies and their correlation (the so-called Tjon line \cite{Tjon}).

It is known \cite{Poly} that when any PET, DET-PET in particular, is applied to  $NN$ interaction,
the binding energy of a three-body (or heavier) system  can be restored by additional 
three-nucleon $NNN$ (or higher-order) interaction(s). 
Our initial $^3$H applications reveal the residual role of the $NNN$
interaction for the ground state energy and how that role changes with the DET-PET selected.  
Similarly, our initial $^4$He applications reveal the residual roles for the combined $NNN$ and 
$NNNN$ interactions on the ground state energy.  Given the numerical challenges of treating 
$NNN$, $NNNN$, etc., interactions in many-body applications, 
it is natural to try to minimize their effects.  In this context, 
DET-PETs are a potentially useful tool in future searches for an $NN$ interaction 
consistent with many-body data.

\section{DET-PET transformation}

Two types of PETs are known in scattering theory: local PETs \cite{Newton} that transform a local 
potential into another local potential and nonlocal PETs \cite{Coester} which generate nonlocal
potential terms. The local PETs always result in some modification of bound state wave
functions \cite{Newton,ShiSid}. Therefore we focus the discussion here on nonlocal PETs.

The Schr\"odinger equation 
\begin{equation}
H\,|\Psi_E\rangle=E\,|\Psi_E\rangle
\label{Schr}
\end{equation}
describes a relative motion in two-body quantum system. The state 
%wave function 
$|\Psi_E\rangle$ can be
expanded in infinite series of $\mathscr L^2$ 
%functions
states $|a_n\rangle$, 
\begin{equation}
|\Psi_E\rangle=\sum_{n=0}^{\infty} c_n(E)\,  |a_n\rangle. 
\label{ExpanPsi}
\end{equation}
The states
% functions 
$|a_n\rangle$
are supposed to form a complete orthonormalized basis,
\begin{equation}
\label{orthog}
\langle a_i |a_j\rangle=\delta_{ij}.
\end{equation}
Using expansion (\ref{ExpanPsi}) we obtain an infinite set of algebraic equations defining the
expansion coefficients $c_n(E) $,
\begin{equation}
\label{summ}
\sum_{n'=0}^{\infty}(H_{nn'}-\delta_{nn'}E)\,c_{n'}(E)=0,
\end{equation}
where $H_{nn'}=\langle a_{n}|H|a_{n'}\rangle $ are the Hamiltonian matrix elements.

A Hamiltonian $\tilde H$ phase-equivalent to $H$ can be defined through its matrix $[\tilde{H}]$
in the basis $\{|a_n\rangle\}$. This matrix $[\tilde{H}]$ can be obtained from $[H]$, 
the matrix of the Hamiltonian $H$ in the basis $\{|a_n\rangle\}$, by means of a
unitary transformation,
\begin{equation}
\label{PET1}
[\tilde{H}]=[U][H][U^{\dagger}].
\end{equation}
The infinite unitary matrix $[U]$ is supposed to be of the form
\begin{equation}
[U]=[U^0]\oplus [I]=\left[
\begin{array}{cc}
[U^0]& 0 \\
0& [I] 
\end{array}
\right],
\label{U0I}
\end{equation}
where  $[I]$ is an infinite unit matrix and $[U^{0}]$, a non-trivial submatrix of $[U]$, is a finite
matrix mixing only  a few selected basis functions. It is clear that
Hamiltonians $H$ and $\tilde{H}$ have
identical eigenvalue spectra. Their eigenstates          %eigenfunctions, 
$|\tilde{\Psi}_E\rangle$ and $|\Psi_E\rangle$, 
differ by a linear combination of a finite number of $\mathscr{L}^2$ basis states.
%functions. 
Any superposition of a finite number of
$\mathscr{L}^2$ functions must decrease
at large distances. Therefore at
positive energy $E$ associated with scattering, the oscillating asymptotics of wave functions 
$\langle\vec{r}|\tilde{\Psi}_E\rangle$ and $\langle\vec{r}|\Psi_E\rangle$ at large distances 
are the same. In other words, the scattering phase
shifts defined through asymptotic behavior of functions $\langle\vec{r}|\tilde{\Psi}_E\rangle$ 
and $\langle\vec{r}|\Psi_E\rangle$
are also the same, i.\;e. the Hamiltonians $H$ and $\tilde{H}$ are phase-equivalent.

The unitary operator $U^0$ can be written as
\begin{equation}
U^{0}=\sum_{i,j\leqslant N} |a_i\rangle {U}^{0}_{ij}\langle a_j|.
\label{U_0}
\end{equation}
The transformation (\ref{PET1})--(\ref{U_0}) leaves the bound state 
%wave function 
$|d\rangle$
unchanged, i.\;e. becomes a DET-PET, when each of the $\mathscr{L}^{2}$ 
%functions
vectors $|a_i\rangle$
entering the non-trivial submatrix $[U^0]$ of the infinite unitary matrix $[U]$ through 
Eq.\ (\ref{U_0}), is orthogonal to $|d\rangle$,\begin{equation}
\langle a_i|d\rangle=0,\ \ \ \ i\le N.
\label{ONdet}
\end{equation}

At this stage, we assert that we have obtained our DET-PET 
defined through the unitary transformation 
(\ref{PET1})--(\ref{U_0}) with 
%functions
vectors $|a_i\rangle$ fitting the conditions (\ref{orthog}) and 
(\ref{ONdet}). In order to obtain a nonlocal interaction $\tilde{V}$ deuteron-equivalent and phase-equivalent
to the initial interaction $V$, we add to $V$ the two-body relative kinetic energy operator $T$ to
obtain the Hamiltonian $H$,
\begin{equation}
H=T+V,
\label{Hini}
\end{equation}
calculate its matrix $[H]$ in the basis $\{|a_n\rangle\}$, obtain the matrix $[\tilde{H}]$ by
means of DET-PET unitary transformation, and obtain the matrix
\begin{equation}
[\tilde{V}]=[\tilde{H}]-[T] .
\label{tildeV}
\end{equation}
Here $[T]$ is the infinite kinetic energy matrix in the basis $\{|a_n\rangle\}$. The  interaction 
$\tilde{V}$ is defined through its matrix $[\tilde{V}]$ in the basis $\{|a_n\rangle\}$. 

The simplest DET-PET is obtained with arbitrary unitary matrix $[U^0]$ of the rank 2. In this
case, $[U^0]$ is  associated either with a rotation by the angle $\beta$ when 
$\det U^0=+1$ or with a rotation by the angle $\beta$ combined with reflection when 
$\det U^0=-1$. We also need to define the $\mathscr{L}^{2}$ vectors $|a_1\rangle$
and $|a_2\rangle$ in Eq. (\ref{U_0}).

We define here the 
%states
vectors $|a_1\rangle$
and $|a_2\rangle$ as  linear combinations of  oscillator states $| \varphi_i \rangle$,
\begin{equation}
|a_i\rangle=\sum_{i'\leqslant N'} \alpha_i^{i'} | \varphi_{i'} \rangle,
\label{b}
\end{equation}
which fit the orthonormality condition  (\ref{orthog}). We expand the deuteron eigenstate
%wave function
$|d\rangle$ in an infinite series of oscillator states, 
% $\varphi_i$,
\begin{equation}
|d\rangle = \sum_{i=0}^{\infty}d_i | \varphi_i \rangle,
\label{detexp}
\end{equation}
where, generally, all the coefficients  $d_i$ are non-zero,
\begin{equation}
d_i\neq 0.
\label{nonzero}
\end{equation}
Since the vectors  $|a_1\rangle$ and $|a_2\rangle$ should fit Eq. (\ref{ONdet}), 
the expansion (\ref{b})  of each of them
involves at least two different basis states  $| \varphi_i \rangle$ due to
Eq. (\ref{detexp})-(\ref{nonzero}). In this simplest case we have
\begin{subequations}
\label{a}
\begin{gather}
|a_1\rangle= a_1^n\, | \varphi_n \rangle +a_1^m\, | \varphi_m \rangle,
\label{a1} \\
|a_2\rangle= a_2^k\, | \varphi_k \rangle +a_2^l\, | \varphi_l \rangle.
\label{a2}
\end{gather}
\end{subequations} 
The normalization of these vectors requires
\begin{subequations}
\label{norm}
\begin{gather}
\left(a_1^n\right)^2+\left(a_1^m\right)^2=1,
\label{norm1} \\
\left(a_2^k\right)^2+\left(a_2^l\right)^2=1,
\label{norm2}
\end{gather}
\end{subequations} 
while the orthogonality of the vectors $|a_1\rangle$ and $|a_2\rangle$,
\begin{equation}
\langle a_2|a_1\rangle=0,
\label{OR}
\end{equation} 
is guaranteed when these vectors are constructed from different basis states, i.\;e.
all the basis states $| \varphi_n \rangle$, $| \varphi_m \rangle$, 
$| \varphi_k \rangle$, $| \varphi_l \rangle$ entering
Eqs. (\ref{a}) are different. Using expansions  (\ref{b}) and (\ref{detexp}) we
obtain
\begin{subequations}
\label{ONdetexp}
\begin{gather}
a_1^n\,d_n+a_1^m\,d_m=0,
\label{ONdetexp1} \\
a_2^k\,d_k+a_2^l\,d_l=0.
\label{ONdetexp2}
\end{gather}
\end{subequations} 
The solutions of Eqs. (\ref{norm}), (\ref{ONdetexp}) are
%(\ref{norm1}), (\ref{norm2}), (\ref{ONdetexp1}), (\ref{ONdetexp2}) are
\begin{subequations}
\begin{gather}
a_1^n=\frac{d_m}{\sqrt{d_n^2+d_m^2}},
\label{sola1n}\\
a_1^m=-\frac{d_n}{\sqrt{d_n^2+d_m^2}},
\label{sola1m}\\
a_2^k=\frac{d_l}{\sqrt{d_k^2+d_l^2}},
\label{sola2k}\\
a_2^l=-\frac{d_k}{\sqrt{d_k^2+d_l^2}},
\label{sola2l}
\end{gather}
\label{solution}
\end{subequations} 

To define completely the simplest DET-PET discussed above we need to fix the rotation angle 
$\beta$, the sign of $\det U^0$ and the set of 4 oscillator states used to build the
states $|a_1\rangle$ and $|a_2\rangle$. To distinguish various DET-PET types we use
notations like $0s 2s 1s 2d^{\pm}$. In this example, the state $|a_1\rangle$  is a
linear combination of the oscillator states $0s$ and $2s$, the vector $|a_2\rangle$  is a
linear combination of the oscillator states $1s$ and $2d$, and the index $\pm$ corresponds
to the sign of $\det U^0=\pm 1$.

\section{DET-PET properties and manifestation in few-nucleon systems}

In this section, we study modifications of the JISP16 $NN$ interaction \cite{JISP16} induced 
by various DET-PETs. The modifications of a nonlocal interaction can be illustrated by 
modifications of its wave functions. The deuteron wave function is unaffected by DET-PET. 
Therefore we present below the DET-PET induced transformation of 
the JISP16 $np$ scattering wave function in the $sd$ coupled partial wave.

 It is interesting to explore a DET-PET which acts only in a single channel, say, in the 
$s$ channel, and compare it with DET-PETs mixing components of the $s$ and
$d$ channels in different ways.
Therefore vectors $|a_1\rangle$ and $|a_2\rangle$ [see Eqs. (\ref{a})] were constructed as
various superpositions of two low-lying
oscillator states of the $np$ relative motion $0s$, $1s$, $2s$,
$3s$, $0d$ and $1d$ with ${\hbar\Omega=40}$~MeV. For each type of the DET-PET we
investigate the transformations associated with both pure rotation and a
rotation-reflection combination.

\begin{figure*}
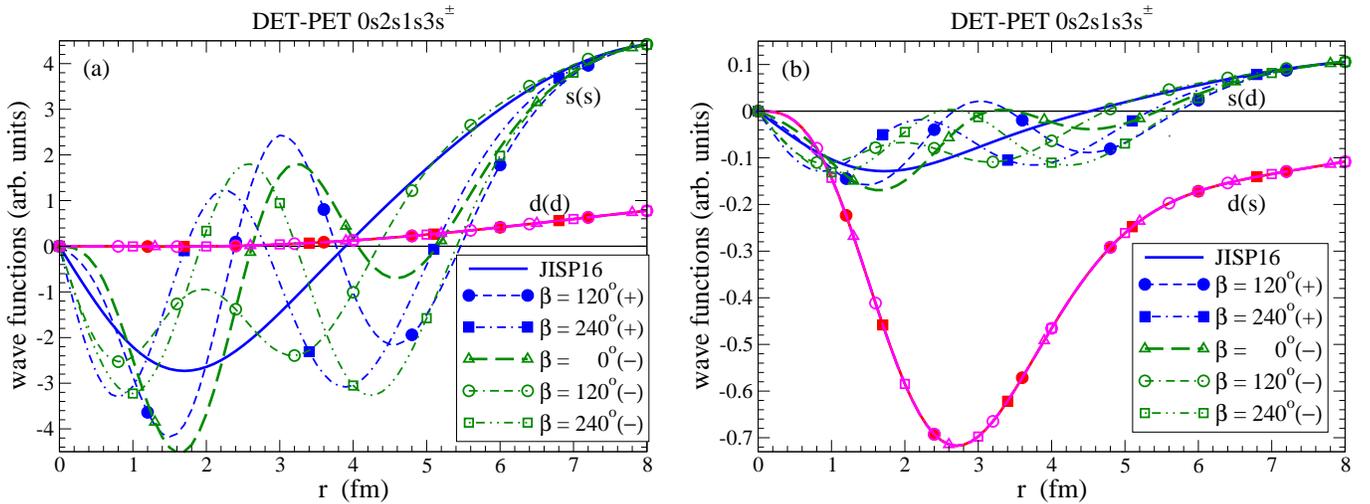

\epsfig{file=0s2s1s3sl.eps,width=\columnwidth}\hfill%
\epsfig{file=0s2s1s3ss.eps,width=1.023\columnwidth}
\caption{(Color online) Large (a) and small (b) components of the $np$ scattering wave 
function at the laboratory energy $E_{\rm lab}=10$~MeV in the $sd$ coupled partial
wave in the $K$-matrix formalism
(see Ref. \cite{ISTP}  for details and nomenclature)
generated by JISP16  and $NN$ interactions obtained from JISP16 by means of
DET-PET $0s2s1s3s^{\pm\!}$. The sign of $\det U^0$ is given in the legends in parenthesis
after the value of rotation angle $\beta$.} 
\label{0s2s1s3s}
\end{figure*}

\begin{figure*}
\epsfig{file=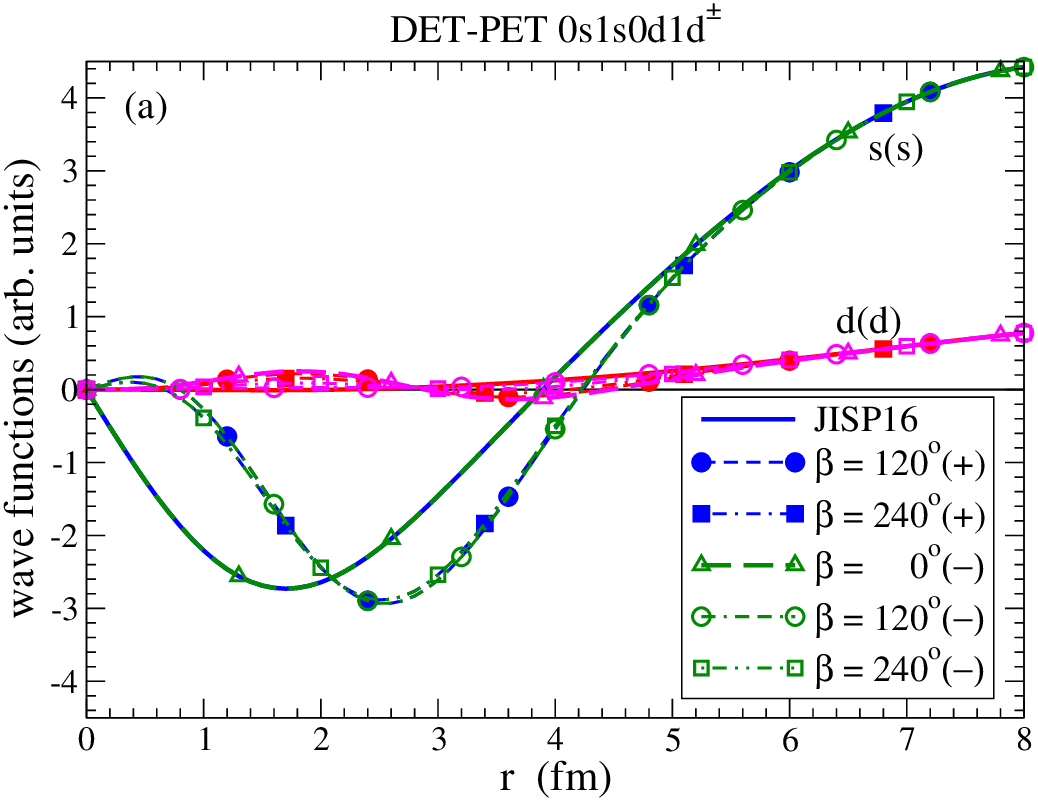,width=\columnwidth}\hfill%
\epsfig{file=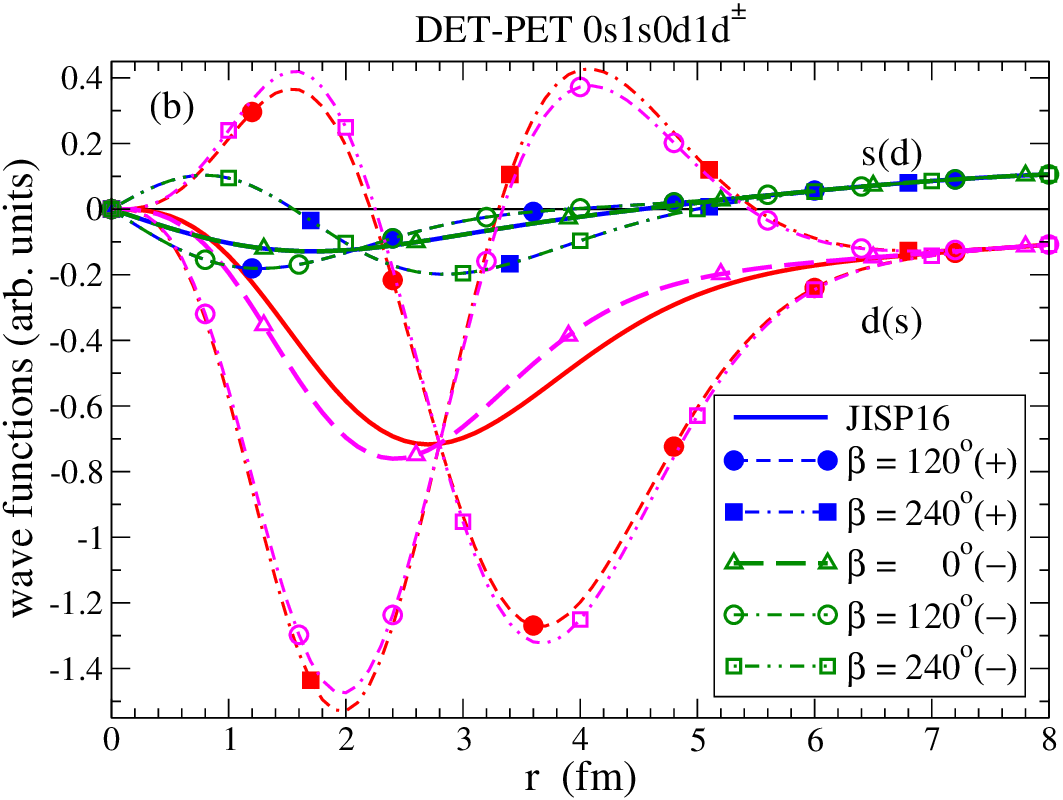,width=1.023\columnwidth}
\caption{(Color online) Same as Fig. \ref{0s2s1s3s} but for DET-PET $0s1s0d1d^{\pm\!}$.}
\label{0s1s0d1d}
\end{figure*}

\begin{figure*}
\epsfig{file=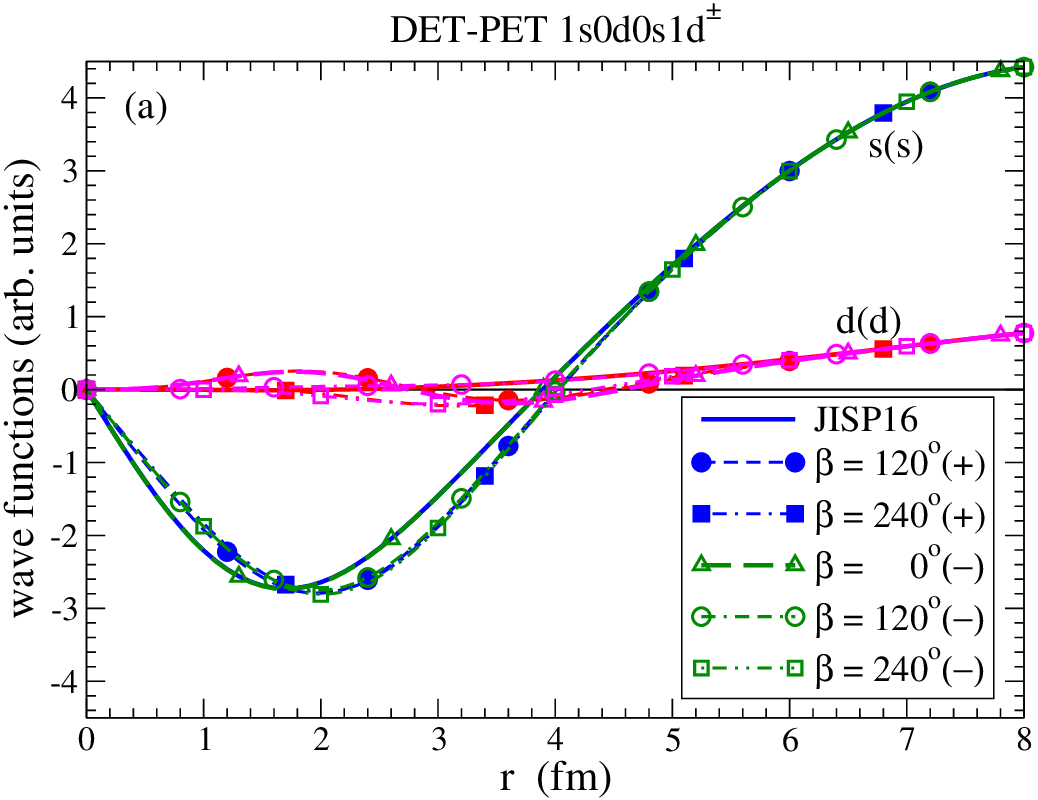,width=\columnwidth}\hfill%
\epsfig{file=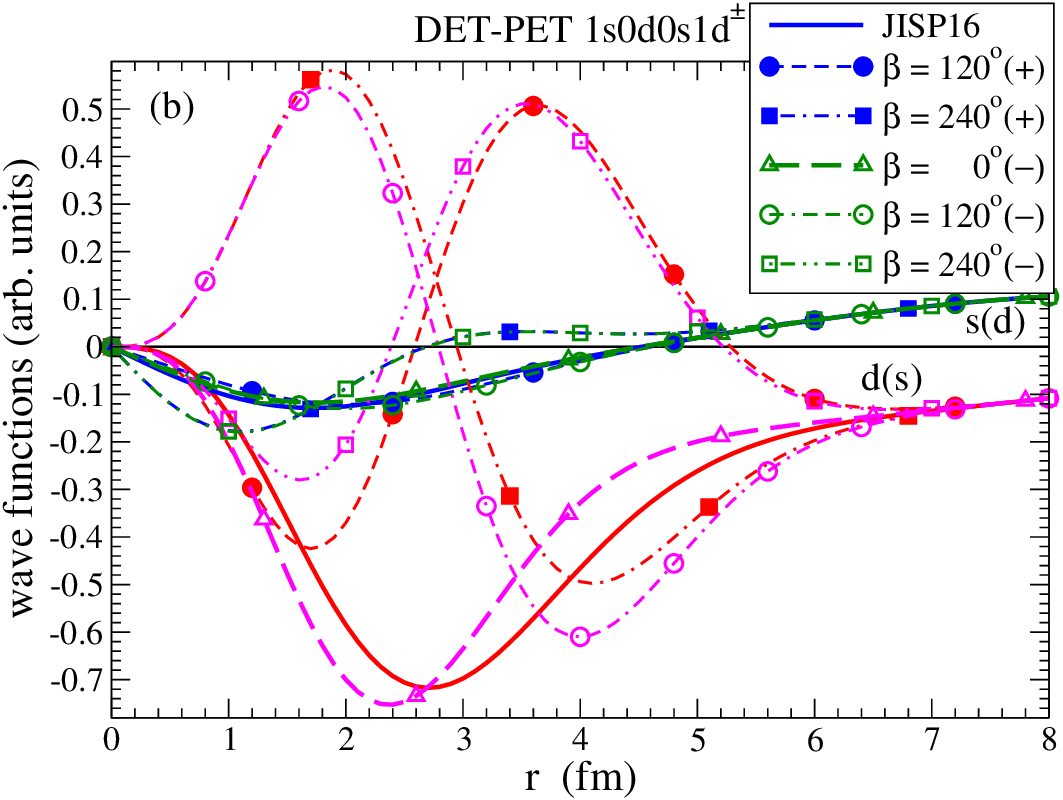,width=1.023\columnwidth}
\caption{(Color online) Same as Fig. \ref{0s2s1s3s} but for DET-PET $1s0d0s1d^{\pm\!}$.}
\label{1s0d0s1d}
\end{figure*}

Plots of the JISP16 $np$ scattering wave functions in the $sd$ coupled partial wave 
at laboratory energy $E_{\rm lab}=10$~MeV and plots for their  $0s2s1s3s^\pm$,
$0s1s0d1d^\pm$ and $1s0d0s1d^\pm$ DET-PET  partners are given in Figs.~\ref{0s2s1s3s},
\ref{0s1s0d1d} and \ref{1s0d0s1d} respectively. We use the $K$-matrix formalism 
(see Ref.~\cite{ISTP}  for details and nomenclature adopted here).
The advantage of the  $K$-matrix formalism is that the
radial wave functions in the scattering domain 
defined according to their  standing wave asymptotics
are real contrary to the more conventional $S$-matrix formalism
with complex radial wave functions which are asymptotically a superposition
of ingoing and outgoing spherical waves.

The DET-PET $0s2s1s3s^{\pm}$ mixes various
$s$ wave components of the wave function that is equivalent to modification of the
central part of the JISP16 interaction in the $s$ wave. This
results in significant changes of the large $s$ wave component
as is seen in Fig.~\ref{0s2s1s3s}. The modification of the small $s$ wave component is
less pronounced. The $d$ wave components, as expected, 
are nearly unaffected by  $0s2s1s3s^{\pm\!}$. 

The DET-PET $0s1s0d1d^{\pm\!}$  explicitly mixes $s$ and $d$ waves; the DET-PET
$1s0d0s1d^{\pm\!}$ also mixes $s$ and $d$ waves but in a different manner.
This corresponds to an essential modification of the tensor component 
of the JISP16 $NN$ interaction. As a result, we observe an essential modification of
small scattering  wave function components which are generated by the tensor
$NN$ interaction as is seen in Figs.~\ref{0s1s0d1d} and \ref{1s0d0s1d}. Modifications
of the large wave function components are much less pronounced. 

We see that DET-PET generates essential modifications of scattering wave functions
without any change of scattering phase shifts and scattering observables (cross
sections, polarization observables, etc.). It
is worth noting here that the deuteron wave function and deuteron observables
(rms radius, quadrupole moment, etc.) are unaffected by DET-PET due to the nature of
this transformation. The alteration of scattering wave functions is an indicator for 
the variation of the off-shell properties of the interaction arising from DET-PET. The
modification of the the $NN$ interaction off-shell should manifest itself in 
properties of many-nucleon systems. Therefore, we investigate  the
DET-PET-induced changes of the $^3$H and
$^4$He binding energies. 

We calculate  $^3$H and $^4$He in the {\em ab initio} No-core Full Configuration 
(NCFC) \cite{NCFC} approach. Within the NCFC approach, we start with the No-core Shell Model 
\cite{NCSM,NCSM2000} calculations using the code 
MFDn \cite{Vary92_MFDn,ACM,SciDAC09,MarisICCS10} 
with a few values of the oscillator frequency $\hbar\Omega$ and in a few basis
spaces characterized by the maximum oscillator quanta $N_{\max}$ 
allowed in the many-body basis  above the minimal configuration. 
Next, we extrapolate the sequence of finite basis space results to the
infinite basis space limit.  This makes it possible to obtain basis
space independent  results for binding energies and to evaluate
their numerical uncertainties.  NCFC suggests two extrapolation methods: a
global extrapolation based on the  calculations in four successive
basis spaces and five $\hbar\Omega$ values in a 10 MeV interval
(extrapolation A), and extrapolation B based on the calculations
at various fixed $\hbar\Omega$ values in three successive basis spaces
and defining the most reliable $\hbar\Omega$ value for the
extrapolation. We present here only the extrapolation~A results based on the NCSM calculations 
with basis spaces up through $N_{\max}=16$. The extrapolations A and B usually provide 
consistent results  \cite{NCFC}, and we checked this consistency for our results  
in a number of cases. The evaluated uncertainties of results for
binding energies presented here are less then 10~keV in most cases; in a few cases, we 
performed the NCSM calculations up to $N_{\max}=18$ to obtain the binding energies with 
uncertainty of about 10~keV.

The binding energies of $^3$H nucleus $E_t$ and of $^4$He nucleus $E_\alpha$ were
calculated with  JISP16 interaction modified by DET-PETs $0s2s1s3s^{\pm}$,
$0s1s0d1d^{\pm}$ and $1s0d0s1d^{\pm}$ varying  angle $\beta$ from $0^{\circ}$ through
$360^\circ$ in steps of $60^\circ$. We observe 
%essential 
variations of $E_t$ and $E_\alpha$ 
due to DET-PETs. In some cases, when the $^3$H and $^4$He binding energies were
close to their maximal or minimal values for a given DET-PET type, we decreased the
step of $\beta$ to investigate the behavior of $E_t$ and $E_\alpha$ around their extremal values
in more detail.

\begin{table}[b]
\caption{Ranges of $^3$H and $^4$He binding energy variations (in MeV) caused by various 
types of  DET-PET in comparison with the binding energies obtained with JISP16 
and their experimental values.}
\label{tminmax}
 \begin{ruledtabular}
 \begin{tabular}{cccc}
 %\multicolumn{2}{c}{$^3$H} & \multicolumn{2}{c}{$^4$He} 
%& \multicolumn{2}{c}{$^3$H} & \multicolumn{2}{c}{$^4$He} 
 %          \\ \cline{1-2} \cline{3-4} \cline{5-6} \cline{7-8}
% $E^b_{\min}$ & $E^b_{\max}$ & $E^b_{\min}$ & $E^b_{\max}$ &
 %$E^b_{\min}$ & $E^b_{\max}$ & $E^b_{\min}$ & $E^b_{\max}$ \\ \cline{1-4} \cline{5-8}
 $^3$H & $^4$He & $^3$H & $^4$He \\ \cline{1-2} \cline{3-4}
%  \multicolumn{4}{c}{$0s2s1s3s^{+}$} &  \multicolumn{4}{c}{$0s2s1s3s^{-}$} \\ 
 \multicolumn{2}{c}{$0s2s1s3s^{+}$} &  \multicolumn{2}{c}{$0s2s1s3s^{-}$} \\ 
 7.21---8.37 & 21.25---28.49 & 7.25---8.35 & 21.46---28.59 \\
  \multicolumn{2}{c}{$0s1s0d1d^{+}$} &  \multicolumn{2}{c}{$0s1s0d1d^{-}$} \\ 
 7.67---8.41 & 23.50---28.83 & 7.68---8.39 & 23.46---28.91 \\
  \multicolumn{2}{c}{$1s0d0s1d^{+}$} &  \multicolumn{2}{c}{$1s0d0s1d^{-}$} \\ 
 7.98---8.64 & 25.79---30.36 & 8.05---8.67 & 26.18---30.41 \\
  \multicolumn{2}{c}{JISP16} &  \multicolumn{2}{c}{Experiment} \\
% \multicolumn{2}{c}{8.369(1)} & \multicolumn{2}{c}{28.299(1)}   
%& \multicolumn{2}{c}{8.482} & \multicolumn{2}{c}{28.296} 
   8.369(1) & 28.299(1) & 8.482 & 28.296
 \end{tabular}
 \end{ruledtabular}
\end{table}

The ranges of $^3$H and $^4$He binding energy variations for each DET-PET type 
are shown in Table~\ref{tminmax}. We see that DET-PETs can cause essential
modification of both $^3$H and $^4$He binding energies. For example, in the case of
the $^4$He nucleus, $E_\alpha$ can be varied by DET-PETs on the interval from 21.25 through
30.41~MeV, i.\;e., the DET-PET
$NN$ interaction can change $E_\alpha$ by more
than 7~MeV from its original value provided by the original JISP16 interaction. In the
case of $^3$H, the range of the DET-PET binding energy variation is 
$7.21\leq E_t\leq 8.67$~MeV, i.\;e., the binding energy can be shifted by more than 1~MeV
from its original JISP16 value.

We study also a correlation of the $^3$H and $^4$He binding energies, the so-called 
Tjon line \cite{Tjon}. The Tjon line is usually studied using results obtained with 
different $NN$ interactions and different combinations of $NN$ and $NNN$
interactions (see, e.\;g.,  Ref. \cite{Nogga}). We note here also an investigation of Jurgenson et al
\cite{Tjon-SRG}  where the $^3$H and $^4$He binding energy correlation   
was studied with $NN$ interactions SRG evolved to various values of momentum parameter 
$\lambda$. An interesting observation mentioned 
by various authors (see, e.\;g., Ref.~\cite{Tjon,Nogga}) is that these results obtained
with different interaction models form nearly a straight line on the plot $E_{\alpha}$
vs $E_t$. Here we study the $E_t$--$E_{\alpha}$ correlation using families of $NN$ potentials
generated by various DET-PET types from the JISP16 interaction,
i.\;e., all $NN$ interactions provide not only algebraically 
identical $NN$ phase shifts but also identical deuteron wave functions that
should give rise to specific $np$
correlations in three- and four-nucleon
systems. 

\begin{figure}
\epsfig{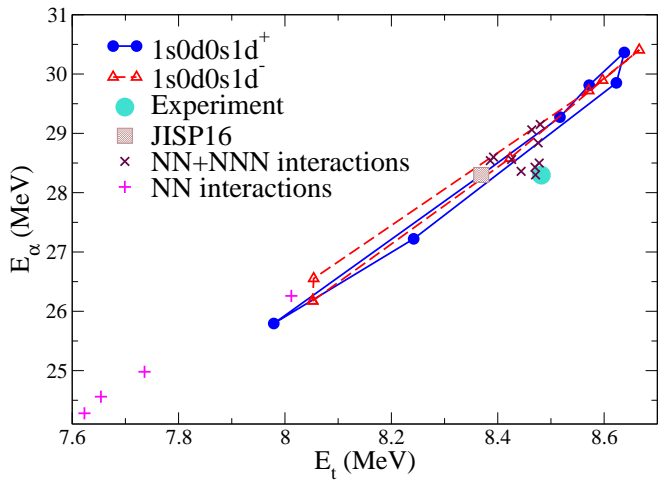}
\caption{(Color online) Tjon line obtained with DET-PET $1s0d0s1d^{\pm}$ in comparison with 
results obtained with various $NN$ and $NN+NNN$ interaction models from 
Refs. \cite{Nogga,N3LO,Pieper}.}
\label{Tjon1s0d0s1d}
\end{figure}

We begin the discussion of the Tjon lines from the results obtained with the
$1s0d0s1d^{\pm}$ DET-PET presented in Fig.~\ref{Tjon1s0d0s1d} where we show also
the results from Refs. \cite{Nogga,N3LO,Pieper} obtained with various modern $NN$ and  
$NN+NNN$ interaction models. It is seen that the DET-PET \mbox{$E_t$--$E_{\alpha}$} correlation 
generally  follows the trend suggested by  other interactions: our results are concentrated
close to the Tjon line connecting the points extracted from Refs. \cite{Nogga,N3LO,Pieper} 
and extend  it to larger $^3$H and $^4$He bindings. We recall here that the
$1s0d0s1d^{\pm}$ DET-PET is associated with  modification of the tensor component
of $NN$ interaction.

\begin{figure}
\epsfig{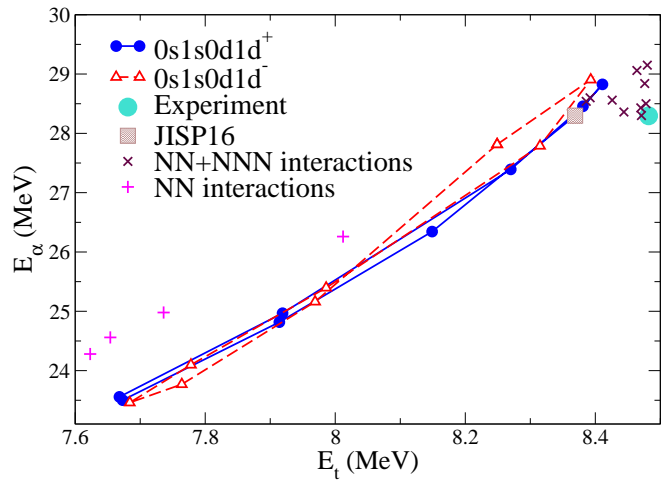}
\caption{(Color online) Same as Fig. \ref{Tjon1s0d0s1d} but for DET-PET $0s1s0d1d^{\pm\!}$.}
\label{Tjon0s1s0d1d}
\end{figure}

Another DET-PET modifying  the tensor component
of the JISP16 $NN$ interaction, is the DET-PET $0s1s0d1d^{\pm\!}$. This DET-PET
results in a very different range of $E_t$ and $E_\alpha$ variations (see Table~\ref{tminmax}).
The DET-PET $^3$H and $^4$He binding energies are also correlated along  a nearly
straight line (see Fig.~\ref{Tjon0s1s0d1d}). However this line has a slope very different
from the slope of the Tjon line obtained with other interaction models. Around the maximal
$^3$H and $^4$He binding energies accessible by this DET-PET, it suggests correlations
consistent with those derived using modern $NN+NNN$ interaction models. However, for
smaller binding energies, this DET-PET suggests much less bound $^4$He at the
same $^3$H bindings as provided by  modern purely two-nucleon interactions.

\begin{figure}
\epsfig{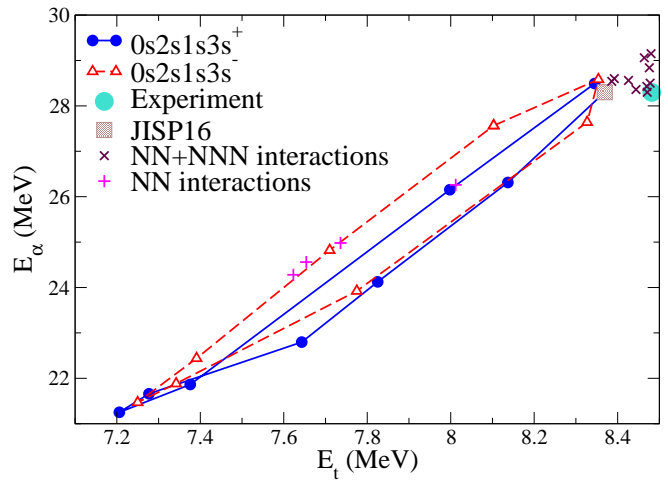}
\caption{(Color online) Same as Fig. \ref{Tjon1s0d0s1d} but for DET-PET $0s2s1s3s^{\pm\!}$.}
\label{Tjon0s2s1s3s}
\end{figure}

The DET-PET $0s2s1s3s^{\pm}$ modifies the central $s$-wave component of the $NN$
interaction. It results in the $^3$H and $^4$He binding energy correlation shown
in Fig.~\ref{Tjon0s2s1s3s}. We see that in this case the results do not concentrate
as tightly around some straight line. That is, they are more spread out 
on the \mbox{$E_t$--$E_{\alpha}$} 
plane. The DET-PET Tjon lines transform into closed-loop curves surrounding elongated 
areas. In the case of the DET-PET $0s2s1s3s^{-}$, the Tjon curve surrounds many points
obtained with various $NN$ interactions.  The DET-PET $0s2s1s3s^{+}$ generates the
Tjon curve shifted down from the Tjon line suggested by other interactions. Both
$0s2s1s3s^{+}$ and $0s2s1s3s^{-}$ DET-PETs essentially extend the range of the
 $^3$H and $^4$He binding energy variations to smaller bindings.

\section{Conclusions}
We have introduced a new type of phase-equivalent transformations, DET-PET, preserving
the deuteron wave function. 
The suggested theory of DET-PET can be easily reformulated to preserve scattering 
wave functions at a given energy instead of the bound state wave function.
We investigated  transformations of the JISP16 $NN$
interaction induced by DET-PETs mixing oscillator components in various combinations. One
of these DET-PETs generates modifications of the central component of the $NN$
interaction, the others modify the tensor $NN$ interaction component. We demonstrated
that DET-PETs are able to modify significantly the $np$ scattering wave functions and hence
the off-shell properties of the $NN$ interaction while the on-shell interaction properties
are preserved. 

DET-PETs impact the binding energies of many-nucleon systems.
We found that the $^3$H and $^4$He binding energies can be significantly
changed by DET-PETs. The investigated DET-PETs modifying
tensor $NN$ interaction, correlate the $^3$H and $^4$He bindings along some lines
that may differ in slope from the Tjon line obtained with modern $NN$ and $NNN$
interactions. The DET-PET $0s2s1s3s^{\pm}$ modifying the central $s$ wave
$NN$ interaction, weakens
the \mbox{$E_t$--$E_{\alpha}$}  correlation spreading the results on the 
\mbox{$E_t$--$E_{\alpha}$}  plane.

It would be interesting to study DET-PET manifestations in binding energies and other 
observables of heavier nuclei. We speculate that DET-PET can be helpful in the further
development of JISP-like $NN$ interactions. 

DET-PETs can be also used to design an
interesting approach to effective interactions. In particular, DET-PETs can be applied to
a modern $NN$ realistic interaction to reduce their high momentum components
(and hence to improve the convergence of
{\em ab initio} calculations). At the same time,  we are preserving the deuteron wave function
and $NN$ correlations in other partial waves at some energy by using the extension of
DET-PET to preserve the scattering wave function. Such an effective interaction can be
very interesting for many-body nuclear applications.

It is also possible to extend the DET-PET concept to $NNN$ interactions.  The corresponding
set of transformations would then involve changing the interior part of the $NNN$ wave function in 
such a manner as to preserve the $NNN$ ground state wave function and to preserve the 
asymptotic $NNN$ wave function.  This could provide a useful tool to explore the off-shell freedoms 
available in $NNN$ interactions without requiring repeated fits to the $NNN$ bound state 
properties.

%\begin{acknowledgments}
%\ acknowledgments
\mbox{}

We are thankful to Peter Sauer for valuable discussions.
This work was supported by the US DOE
Grants DE-FC02-09ER41582 and DE-FG02-87ER40371 and Ministry of Education
and Science of the  Russian Federation under the Contract P521.
%\end{acknowledgments}

\mbox{}

\mbox{}


\begin{thebibliography}{99}
\bibitem{Newton} R. G. Newton, {\em Scattering theory of waves and
particles,  2nd. ed.} (Springer-Verlag, New York, 1982).

\bibitem{SUSY-review} F. Cooper, A. Khare, and U. Sukhatme, Phys. Rep. {\bf 251},  267 (1995).

\bibitem{SUSY-inversereview} D. Baye and J.-M. Sparenberg, J. Phys. A {\bf 37}, 10223 (1994).

\bibitem{ShiSid} A. M. Shirokov  and V. N.  Sidorenko,  Yad. Fiz. {\bf  63}, 2085 (2000)
[Phys. At. Nucl. 63, 1993 (2000)].

\bibitem{Coester} F. Coester, S. Cohen, B. Day, and C. M. Vincent,  Phys. Rev. C {\bf 1}, 769 (1970).

\bibitem{PHT} Yu.~A.~Lurie and A.~M.~Shirokov,  Izv. Ros. Akad. Nauk,
Ser. Fiz. {\bf 61}, 2121 (1997) [Bull.
Rus. Acad. Sci., Phys. Ser. {\bf  61}, 1665 (1997)].

\bibitem{LurAnn} Yu.~A.~Lurie and A.~M.~Shirokov, Ann. Phys. (NY)  {\bf 312}, 284 (2004);
in {\itshape The $J$-Matrix Method. Developments and
     Applications}, edited by A. D. Alhaidari, H. A. Yamani, E. J. Heller, and
     M. S. Abdelmonem (Springer, 2008), 183.



 \bibitem{JISP6} A. M.  Shirokov,  J. P. Vary,  A. I. Mazur, S. A. Zaytsev,  and
T. A. Weber,  Phys. Lett. B {\bf 621}, 96 (2005);  
 J. Phys. G {\bf 31},  S1283 (2005).  

\bibitem{JISP16}  A. M. Shirokov,  J. P. Vary,  A. I. Mazur, and
T. A. Weber,  Phys. Lett. B {\bf 644}, 33 (2007).
 
 \bibitem{JISP16_web} A Fortran code for the JISP16 interaction matrix elements
is available at http:/\!/nuclear.physics.iastate.edu.
 
\bibitem{ISTP}   A. M. Shirokov,  A. I. Mazur,   S. A. Zaytsev,  J. P. Vary,  and
T. A. Weber,   Phys. Rev. C {\bf 70},  044005 (2004);
in {\itshape The $J$-Matrix Method. Developments and
     Applications}, edited by A. D. Alhaidari, H. A. Yamani, E. J. Heller, and
     M. S. Abdelmonem (Springer, 2008), 219.

\bibitem{LIT-JISP} N. Barnea, W. Leidemann, and G. Orlandini, Phys. Rev. C {\bf 74}, 034003 (2006);
W. Leidemann, Nucl. Phys. A {\bf 790}, 24 (2007); G. Orlandini, S. Bacca, N. Barnea, and 
W.~Leidemann, Nucl. Phys. A {\bf 790}, 368 (2007).

\bibitem{Slaus2007} I. \v{S}laus, Nucl. Phys. A {\bf 790}, 199 (2007).

\bibitem{YaF2008} A. M. Shirokov, J. P. Vary, A. I. Mazur, and T. A. Weber, 
Yad. Fiz. {\bf 71}, 1260 (2008) [Phys. At. Nucl. {\bf 71}, 1232 (2008)].

\bibitem{Rila} A. M. Shirokov, J. P. Vary, and P. Maris, in {\em Proc. of the 27th
Int. Workshop on Nucl. Theory, Rila Mountains, Bulgaria, 23--28 June,
2008}, edited by S.Dimitrova (%Institute for Nuclear Research and Nuclear Energy, 
Bulgarian Academy of Science, 2008),  205 [arXiv:0810.1014 (2008)].

\bibitem{JMPE2008} J. P. Vary, P. Maris, and  A. Shirokov,  Int. J. Mod. Phys. E {\bf 17}, Suppl.~1, 109 (2008). 

\bibitem{NCFC} P. Maris, J. P. Vary, and A. M. Shirokov, 
Phys. Rev. C {\bf 79}, 014308 (2009).


\bibitem{VainBarnGaz} S. Vaintraub, N. Barnea, and D. Gazit, Phys. Rev. C {\bf 79}, 065501 (2009).

\bibitem{fbBonn} A. M. Shirokov, V. A. Kulikov, P. Maris, A. I. Mazur, E. A. Mazur, and 
J. P. Vary, EPJ Web of Conf. {\bf 3}, 05015 (2010); in {\em Proc. 3rd Int. Conf. Current
Problems in Nucl. Phys. and Atomic 
Energy  (NPAE--Kyiv2010),  June 7--12, 2010, Kyiv,
Ukraine} (Kyiv, 2011), Part I,  321 [arXiv:1009.2993 (2010)].

\bibitem{OrlBarnLeid} N. Barnea, W. Leidemann, and G. Orlandini, Phys. Rev. C {\bf 81}, 064001 (2010); G. Orlandini, N. Barnea, and W.~Leidemann, J. Phys. Conf. Ser. {\bf 312}, 092049 (2011).

\bibitem{Izv2011} A. M. Shirokov, J. P. Vary, V. A. Kulikov, P. Maris, A.~I.~Mazur, and  E. A. Mazur,
 Izv. Ros. Akad. Nauk, Ser. Fiz. {\bf 75}, 499 (2011)  [Bul. Rus. Acad. Sci.: Phys. {\bf 75}, 463 (2011)]. 


\bibitem{14F}  P. Maris,  A. M. Shirokov, and J. P. Vary,  Phys. Rev. C {\bf 81},  
  021301 (2010).

\bibitem{14Fexp} V. Z. Goldberg,  B. T. Roeder, G. V. Rogachev,  G.~G.~Chubarian,  E. D. Johnson,  C. Fu,  A. A. Alharbi, M. L. Avila, A. Banu, M. McCleskey,  J. P. Mitchell,  E.~Simmons,  G. Tabacaru, L. Trache, and R. E. Tribble, Phys. Lett. B {\bf 692}, 307 (2010).

\bibitem{Tjon} J. A.  Tjon,  Phys. Lett. B {\bf 56}, 217 (1975).

\bibitem{Poly} W. N. Polyzou and W. Gl\"ockle, Few-Body Syst. {\bf 9},
97 (1990).

\bibitem{NCSM} D.~C. Zheng, J. P. Vary, and B.~R.~Barrett, Phys. Rev.
C {\bf 50}, 2841 (1994); D.~C.~Zheng,  B.~R.~Barrett, J.~P.~Vary,
W.~C.~Haxton, and C.~L.~Song, Phys. Rev. C {\bf 52}, 2488 (1995).


\bibitem{NCSM2000}  P.~Navr\'atil, J.~P.~Vary, and B.~R.~Barrett,
Phys. Rev. Lett. {\bf 84}, 5728 (2000);
Phys. Rev. C {\bf 62}, 054311 (2000).

\bibitem{Vary92_MFDn}
J.~P.~Vary, {\em The Many-Fermion-Dynamics Shell-Model Code}, Iowa State University,
1992 (unpublished); J.~P.~Vary and D.~C.~Zheng, {\it ibid}, 1994  (unpublished);
demonstration runs can be performed through http:/\!/nuclear.physics.iastate.edu/mfd.php.

\bibitem{ACM}P.~Sternberg, E.~G.~Ng, C.~Yang, P.~Maris, J. P.~Vary, M.~Sosonkina, and H.~V.~Le,
%"Accelerating Configuration Interaction Calculations for Nuclear Structure", 
in {\em Proc. of the 2008 ACM/IEEE Conf. on Supercomputing} (Austin, TX, November 15-21,
2008),
%Conference on High Performance Networking and Computing.
(IEEE Press, Piscataway, NJ), p. 1.
%, http://doi.acm.org/10.1145/1413370.1413386.

\bibitem{SciDAC09} J. P. Vary,  P. Maris, E. Ng, C. Yang, and M. Sosonkina,
%"Ab initio nuclear structure - the large sparse matrix eigenvalue problem,"
J.\ Phys.\ Conf.\ Ser.\  {\bf 180}, 012083 (2009).

\bibitem{MarisICCS10} P. Maris, M. Sosonkina,  J. P. Vary,  E. Ng and C. Yang,
 %title = {Scaling of ab-initio nuclear physics calculations on
 %         multicore computer architectures},
% Procedia Computer Science 
Proc. Comp. Sci. {\bf 1}, 97 (2010).
%doi = "10.1016/j.procs.2010.04.012",
%adsurl = "http://www.sciencedirect.com/science/article/pii/S187705091000013X"



\bibitem{Nogga}  A. Nogga, H. Kamada, and W.~Gl\"ockle, Phys. Rev. Lett. {\bf 85}, 944 (2000). 

\bibitem{Tjon-SRG} E. D. Jurgenson, P. Navr\'atil, and R. J. Furnstahl, Phys. Rev. Lett. {\bf 103},
082501 (2009).

\bibitem{N3LO} A. Nogga, P. Navr\'atil, B.~R.~Barrett, and J.~P.~Vary,  Phys. Rev. C {\bf 73}, 
064002 (2006).

\bibitem{Pieper} I. Brida, S. C. Pieper, and R.~B.~Wiringa,  Phys. Rev. C  {\bf 84}, 024319 (2011). 


\end{thebibliography}
\end{document}